# Interlayer Coupling and Ultrafast Hot Electron Transfer Dynamics in Metallic VSe$_2$/Graphene van der Waals Heterostructures

Tae Gwan Park[1,†], Byoung Ki Choi[2,†], Junho Park[1], Jungdae Kim[3], Young Jun Chang[2,4,*] and Fabian Rotermund[1,*]

[1]Department of Physics, Korea Advanced Institute of Science and Technology (KAIST), Daejeon 34141, Republic of Korea

[2]Department of Physics, University of Seoul, Seoul 02504, Republic of Korea

[3]Department of Physics, University of Ulsan, Ulsan 44610, Republic of Korea

[4]Department of Smart Cities, University of Seoul, Seoul 02504, Republic of Korea

**ABSTRACT**

Atomically thin vanadium diselenide (VSe$_2$) is a two-dimensional transition metal dichalcogenide exhibiting attractive properties due to its metallic 1T-phase. With the recent development of methods to manufacture high-quality monolayer VSe$_2$ on van der Waals materials, the outstanding properties of VSe$_2$-based heterostructures have been widely studied for diverse applications. Dimensional reduction and interlayer coupling with a van der Waals substrate lead to its distinguishable characteristics from its bulk counterparts. However, only a few fundamental studies have investigated the interlayer coupling effects and hot electron transfer dynamics in VSe$_2$ heterostructures. In this work, we reveal ultrafast and efficient interlayer hot electron transfer and interlayer coupling effects in VSe$_2$/graphene heterostructures. Femtosecond time-resolved reflectivity measurements showed that hot electrons in VSe$_2$ were transferred to graphene within a 100-fs timescale with high efficiency. Besides, coherent acoustic phonon dynamics indicated interlayer coupling in VSe$_2$/graphene heterostructures and efficient thermal energy transfer to three-dimensional substrates. Our results provide valuable insights into the intriguing properties of metallic transition metal dichalcogenide heterostructures and motivate designing optoelectronic and photonic devices with tailored properties.

**INTRODUCTION**

Two-dimensional transition metal dichalcogenides (TMDs) have been extensively studied because of their exceptional characteristics, including tunable bandgaps, phase transitions, and superconducting behavior.[1-3] For example, vanadium diselenide ($VSe_2$) in its metallic 1T-phase exhibits many attractive properties, such as charge-density-waves (CDWs) and magnetic ordering.[4] Recently, uniform monolayer $VSe_2$ has been produced, and its remarkable physical properties unlike those in the bulk phase have been extensively investigated, including strong room-temperature ferromagnetism[5,6] and distinct CDW-modulated structures.[7-9] Motivated by the reported exotic characteristics at the single-layer limit, studies on the effects of dimensional confinement have revealed unusual features of monolayer $VSe_2$. Applications have also been recently introduced that exploit the metallic properties of the high 2D electrical conductivity in $VSe_2$,[10] such as superior potassium-ion storage and flexible in-plane solid-state supercapacitors.[11,12]

In most previous investigations, flat monolayer (1L) $VSe_2$ has been produced using van der Waals substrates, like graphene (Gr) and graphite.[5,7-9] The notable characteristics of 1L $VSe_2$ can be attributed to not only dimensional confinement but also interlayer coupling between $VSe_2$ and Gr. Density functional theory (DFT) calculations have predicted the hole doping of graphene *via* charge transfer from Gr to $VSe_2$ due to weak van der Waals coupling.[13] Also, metal-to-insulator transition resulting from charge ordering at the $VSe_2$–Gr heterointerface has been observed using scanning tunneling microscopy (STM) and angle-resolved photoemission spectroscopy (ARPES).[7] These results show that interlayer coupling between $VSe_2$ and Gr fundamentally influences the properties of low-dimensional layered materials. However, extensive and systematic studies on the interlayer coupling effect in VSe$_2$/van der Waals materials and related heterointerfaces are still lacking.

Interlayer coupling effects have been frequently evidenced by interlayer charge/energy transfer. Optical methods such as photoluminescence (PL) and ultrafast spectroscopy have been employed to investigate interlayer charge transfer in various heterostructures, including semiconducting TMD-based van der Waals heterostructures (vdWHs) such as AB$_2$/AB$_2$[14,15] or AB$_2$/Gr[16-18] (A: Mo, W and B: S, Se). Interestingly, PL quenching in semiconducting TMD/Gr heterostructures has provided intuitive evidence of such interlayer charge transfer.[16-18] Studies using transient absorption spectroscopy (TAS) with an ultrabroadband probe have provided comprehensive pictures of interlayer charge transfer by monitoring carrier dynamics in frequency and time domains.[14-16,18,19] However, PL measurements and TAS analysis are relatively challenging for VSe$_2$/Gr heterostructures because of the absence of optical transition since no bandgap exists in metallic electronic structures. However, time-resolved spectroscopy can be used since ultrafast carrier dynamics in VSe$_2$ and Gr are distinguishable.

In this work, we investigated carrier dynamics in 1L VSe$_2$ and ultrafast interlayer hot electron transfer dynamics in monolayer (1L) and multilayer (MtL) VSe$_2$/Gr heterostructures by femtosecond time-resolved reflectivity measurements in a two-color pump and probe scheme. The experimental results directly showed the influence of interlayer coupling in VSe$_2$/Gr heterostructures. We observed interlayer coupling and ultrafast interlayer hot electron transfer between Gr and VSe$_2$. The photo-injected hot electrons in VSe$_2$ were instantaneously transferred to Gr with a timescale of 100 fs and efficiency of 96% due to the combination of the unusually slow dynamics of VSe$_2$ and ultrafast dynamics in Gr. In MtL VSe$_2$/Gr heterostructures, we observed slower hot electron transfer to Gr, occurring on a 1-ps timescale. We verified energy transfer from VSe$_2$/Gr heterostructures to the SiC substrate *via* the capture of longitudinal out-of-plane coherent acoustic phonons in SiC.

**RESULTS AND DISCUSSION**

**Preparation and characterization of 1L VSe$_2$/Gr heterostructures**

VSe$_2$ samples were grown by molecular beam epitaxy (MBE) on epitaxial Gr, which was first thermally grown on the SiC substrate by epitaxy (Figure 1a). 1L VSe$_2$ was directly grown on a bare 6H-SiC substrate. 1L and MtL VSe$_2$/Gr heterostructures were fabricated by utilizing the same method on Gr. MBE-grown samples were centimeter-size with a uniform crystalline 1T-phase of VSe$_2$.[7] About 1.5-nm-thick Pd was additionally deposited as a capping layer on top of VSe$_2$ for protection from air and humidity. Detailed information about sample growth and characteristics is provided in the Methods.

We performed Raman spectroscopy to characterize the structural and electronic properties of Gr. The Raman spectra in Figure 1b were obtained by the subtraction of the reference spectrum from the SiC substrate.[20] For Gr, the G and 2D peaks were observed at 1594 and 2732 cm$^{-1}$, respectively. The sharp 2D peak with a full-width at half maximum (FWHM) of 53 cm$^{-1}$ shows single-layer characteristics.[20,21] The Gr grown on *n*-type SiC substrate is typically *n*-doped.[22-24] As Raman frequencies are highly sensitive to the doping level of Gr, we estimated the charge carrier concentration of Gr by carefully reviewing the G peak frequency.[24,25] The observed blue shift of the G peak as much as 11 cm$^{-1}$ with respect to neutral Gr (1583 cm$^{-1}$)[25] indicates that our Gr was doped with an electron concentration of ~1 × 10$^{13}$ cm$^{-2}$, which corresponds to a Fermi level of 0.3 eV.

For 1L VSe$_2$/Gr, two significant Raman peaks of Gr were observed at 1587 and 2713 cm$^{-1}$ (Figure 1b). The slightly redshifted G peak (about 7 cm$^{-1}$) indicated that the doping level of Gr decreased by charge transfer to VSe$_2$ for Fermi level alignment during the thermal equilibrium. The electron concentration of Gr in VSe$_2$/Gr was estimated to be about $0.1 \times 10^{13}$ cm$^{-2}$, corresponding to a Fermi level below 0.1 eV, close to the charge neutral point. This *p*-doping of Gr by the charge transfer to VSe$_2$ was predicted in a previous DFT study,[13] which agrees well with our measurements. Note that the Fermi level of Gr in 1L VSe$_2$/Gr was located at an almost neutral point (Figure 1c). The redshift of the 2D peak (about 15 cm$^{-1}$) in 1L VSe$_2$/Gr can be attributed to the *p*-doping and strain induced by VSe$_2$ layer.[20,21] The D peak at 1360 cm$^{-2}$ observed in both Gr and 1L VSe$_2$/Gr is attributed to domain boundaries and defects in Gr.

In Figure 1d-e, the ARPES results of 1L VSe$_2$ on 1L Gr are illustrated. Figure 1d shows valence band dispersion along the Γ–M–K direction. The dispersions of the hole pocket near the Γ point, and in particular, the dispersion at the middle of the M–K direction touching the Fermi level agreed well with the theoretical calculation of 1L VSe$_2$ for the metallic 1T-phase.[26] In Figure 1e, the single linear dispersion of Gr at the K point of Gr indicates that our Gr under VSe$_2$ is single layer Gr.[27] The charge neutral point of Gr under 1L VSe$_2$ is close to the Fermi level, which is consistent with the Raman results.

**Ultrafast interlayer hot electron transfer in 1L VSe$_2$/Gr heterostructures**

To study the interlayer coupling between Gr and VSe$_2$, we performed time-resolved two-color pump-probe spectroscopy with an 830-nm (1.5 eV, ~100 fs) pump and 1120–1600 nm (1.11– 0.77 eV, ~150 fs) probe pulses. The pump pulses were used to generate hot electrons in 1L VSe$_2$ and Gr. The near-infrared (NIR) probe band was selected to avoid Fermi level shift due to the presence of 1L VSe$_2$, which was sensitively observable in the mid-IR/THz region.[22,28]

We firstly verified that the influences of the Pd capping layer and SiC substrate were negligible (see Note 1 in the Supporting Information). Figure 2 shows the time-resolved transient reflectivity (TR) results for Gr, 1L VSe$_2$, and 1L VSe$_2$/Gr heterostructures excited at a pump fluence of 30 µJ/cm$^2$. The TR measurements were performed in the linear-response regime to avoid additional influences, such as Fermi level changes from the nonlinear response of Gr[29] (see Note 2 in the Supporting Information).

The probe-energy-dependent TR signals in the 2D plot (Figures 2a-c) showed significant changes in the carrier dynamics of VSe$_2$ upon forming heterostructures with Gr. The vertical cut of the 2D plot indicates the spectral response under pump excitation (Figure 2d). The horizontal cut of the 2D plot indicates the dynamic response (Figure 2e). Gr and 1L VSe$_2$/Gr exhibited photo-bleaching (PB, negative sign in TR), and 1L VSe$_2$ exhibited photoinduced absorption (PIA, positive sign in TR) in the entire probe energy region. The spectral response of Gr originates from Pauli blocking in the linear Dirac band structure. The excited and thermalized hot electrons and holes prevent additional probe photon absorption. The monotonic reduction in the TR signal at lower probe energies is attributed to the density of states for Gr. The dynamic response of thermalization with a fast-rising signal (~0.2 ps) and cooling dynamics with decay within ~1 ps agree will with previous studies.[30,31] It corresponds to ultrafast relaxation due to the strong coupling between electrons and high-energy optical phonons.

In 1L VSe$_2$ without Gr, the spectral and dynamic response was highly distinguishable compared to those of Gr, as shown in Figures 2d and 2e. The PIA response in 1L VSe$_2$ was mainly attributed to the band structure of 1T-VSe$_2$. According to the theoretical calculation of the band structure of VSe$_2$,[26] the energy difference between the valence band maximum and Fermi level is about 1.6 eV. Subsequently, an intraband transition in the valence band is possible under 1.5 eV excitation. The energy difference between the valence band and conduction band is less than 0.3 eV. Consequently, excited electrons can readily absorb probe photons without a momentum shift, and the interband transition occurs in the probe energy band, *i.e.*, PIA. The spectral response monotonically decreases as the probe energy decreases. The carrier dynamics of 1L VSe$_2$ were substantially slow compared to those of Gr (Figures 2e and 2f). The TR signal slowly increased to 2.5 ps (hot electron relaxation, $\tau_{relax}$), then, decayed with two time constants, 16 and 200 ps, respectively, which was considerably slower compared to decay time of Gr. These decay constants were almost independent of the pump fluence (Figure S3a in the Supporting Information). This slow cooling behavior may be caused by two factors: intrinsic weak electron-phonon coupling in VSe$_2$ CDW systems and the relatively low frequency of the highest-energy optical phonon in VSe$_2$ (~90 meV)[32] compared to that in Gr (optical G phonon of ~200 meV).[33]

In 1L VSe$_2$/Gr heterostructures, the results on carrier dynamics showed a significant change, which was closely related to the interlayer hot electron transfer. The spectral and dynamic responses of TR were quite different from those in 1L VSe$_2$ but similar to those in Gr. The PB signal in the spectral TR response was observed throughout the entire probe range, as shown in Figure 2d. The TR signal quickly increased within 0.2 ps and decayed first within a period comparable to that of Gr (Figure 2e). Relatively slow decays were then observed with two additional decay times (6 and 38 ps), which were faster than those of 1L VSe$_2$ (Figure 2f). By comparing the TR signal of 1L VSe$_2$/Gr heterostructures with those of Gr and 1L VSe$_2$, the following three characteristics reveal the interlayer hot electron transfer from VSe$_2$ to Gr (Figure 3a): (1) The PIA signal in 1L VSe$_2$ was completely quenched, and only the PB signal was observed.

(2) The PB signal in Gr was substantially larger in the entire probe band. (3) The decay times of 1L VSe$_2$ became more efficient by Gr, which served as an efficient cooling channel. Figure 3b shows that the fast-rising PB signal was almost the same in Gr and 1L VSe$_2$/Gr, while the PIA signal slowly increased in 1L VSe$_2$ ($\tau_{relax}$ ~2.5 ps). The nearly same rising time in Gr and 1L VSe$_2$/Gr implies that the timescale of the interlayer hot electron transfer ($\tau_{transfer}$) was shorter than the temporal resolution (about 100 fs). Based on the timescales of the hot electron relaxation and interlayer transfer, we estimated the interlayer electron transfer efficiency[34] ($\eta = 1 - \tau_{transfer}/\tau_{relax}$) to be 96%.

Ultrafast spectroscopic measurements further showed the interlayer coupling effect in 1L VSe$_2$/Gr. The injected hot electrons in 1L VSe$_2$ were transferred to Gr across the 2D interface within a timescale of 100 fs. The timescale of interlayer electron transfer can be tens of femtoseconds, as previously investigated for other semiconducting TMD-based heterostructures by the phonon-mediated interlayer charge transfer model,[14,15,35,36] indicating that the interlayer transfer efficiency might be higher than the estimated value. Consequently, highly efficient interlayer electron transfer from 1L VSe$_2$ provides great potential for the photonic and optoelectronic applications of Gr, and the linear and nonlinear photoresponsivity of Gr can be much enhanced by VSe$_2$. Semiconducting TMDs have been applied to Gr-based devices by charge transfer.[16-18,37-42] However, the intrinsic properties of semiconductors limit the applicable spectral range above the bandgap (mostly in the visible range). In this regard, 1L VSe$_2$ is a promising candidate for ultrabroadband applications in the infrared spectral range due to the metallic character. The MBE-grown 1L VSe$_2$ on thermally grown epitaxial Gr on SiC showed excellent interlayer coupling and is suitable for large-scale production. Notably, the efficiency of the interlayer hot electron transfer was high, although single-layer Gr contained domain boundaries and defects (D-peak in Figure 1b), owing to the slow hot electron relaxation in 1L VSe$_2$. In addition, the thickness of Gr can be controlled by the growth temperature and rate.[20] We also observed the same interlayer electron transfer results for 1L VSe$_2$ on multilayer graphene (MLG) (see Note 3 in the Supporting Information).

**Ultrafast hot electron dynamics in MtL VSe$_2$/Gr heterostructures**

We further studied interlayer hot electron transfer in multilayer (MtL) VSe$_2$/Gr heterostructures, motivated by recent studies on MtL VSe$_2$ nanosheets[11,12] and heterostructures with Gr.[43,44] The interlayer hot electron transfer from VSe$_2$ to Gr instantaneously breaks the population symmetry of the injected hot electrons in MtL VSe$_2$. Consequently, we verified the interlayer hot electron transfer characteristics in MtL VSe$_2$/Gr heterostructures. Figure 4 shows the measured carrier dynamics in MtL (3L, 5L, and 7L) VSe$_2$/Gr heterostructures.

In our TR measurements with a 1.5-eV pump and 1.0-eV probe, the TR signal in MtL VSe$_2$/Gr was a mixture of the enhanced PB signal (negative TR) achieved in Gr by hot electron transfer from VSe$_2$ and the PIA signal (positive TR) of VSe$_2$ itself, as shown in Figure 4a. The negative PB signal rose and decayed quickly as for Gr, and the positive PIA response gradually emerged around 1 ps and increased with the number of VSe$_2$ layers. The distinguishable dynamic response of the fast PB and slow PIA signals gives useful information to analyze hot electron transfer in MtL VSe$_2$/Gr.

The PIA signal in MtL VSe$_2$/Gr was considerably smaller than that in 1L VSe$_2$. For example, the TR value of the PIA signal in 7L VSe$_2$/Gr was about 60% smaller than that in 1L VSe$_2$ without Gr. This result contradicts the assumption that MtL VSe$_2$ was isolated, indicating that hot electrons in MtL VSe$_2$ transferred toward Gr at a timescale of 1 ps (faster than the hot electron relaxation time, ~2.5 ps in 1L VSe$_2$). However, positive PIA signals were observed in only 5L and 7L VSe$_2$/Gr heterostructures. It is unclear whether the origin of this difference is the timescale of hot electron transfer between VSe$_2$–VSe$_2$, which was quite slow compared to VSe$_2$-Gr (~100 fs), or the stronger PIA signal, which increases with the number of VSe$_2$ layers. In the first scenario, the dynamic response at each MtL VSe$_2$/Gr varies significantly as the number of VSe$_2$ layers increases, and consequently, MtL VSe$_2$ can be treated as independent monolayers, as previously studied in incommensurate heterostructures.[45] Conversely, in the second scenario, the dynamic response does not change significantly although the number of VSe$_2$ layers increases. This means that the MtL VSe$_2$ can be considered as bulk for strong interlayer coupling and hybridization between VSe$_2$ layers.[2,16,46]

To confirm this, we performed the same measurements at various pump fluences. Figure 4b shows the pump-fluence-dependent TR signal in 3L VSe$_2$/Gr, emphasizing the signal contributed from MtL VSe$_2$. Near 1 ps after the fast decay in Gr, the PB signal increased, mainly contributed by hot carriers in Gr, until reaching a pump fluence of 30 μJ/cm$^2$. When the pump fluence exceeded 30 μJ/cm$^2$, the trend reversed and the PIA signal, mainly contributed by hot electrons in VSe$_2$, appeared at around 45 μJ/cm$^2$ (Figure 4c). Note that the dynamic response of 3L VSe$_2$/Gr at a higher fluence of 65 μJ/cm$^2$ was almost the same as that of 5L VSe$_2$/Gr at a fluence of 30 μJ/cm$^2$ (Figure 4b), implying that the TR signal we observed in MtL VSe$_2$/Gr follows the second scenario and that VSe$_2$ multilayers were well-stacked by van der Waals epitaxy. In 5L and 7L VSe$_2$/Gr, the zero-crossing was not visible due to the strong PIA signal in thicker VSe$_2$ layers, and the increasing trend of the PIA signal at 1 ps for the pump fluence was proportional to the number of VSe$_2$ layers (Figure 4c).

Figure 3 shows the decay properties of the observed PIA signal (positive TR) in MtL VSe$_2$/Gr heterostructures. Compared to 1L VSe$_2$/Gr, the decay times were slow due to the intrinsic long decay of hot electrons in VSe$_2$ without Gr. However, the shortened decay times compared to that of 1L VSe$_2$ (~200 ps) provide additional evidence on interlayer hot electron transfer from MtL VSe$_2$ to Gr and accelerate the relaxation process. The injected hot electrons in MtL VSe$_2$ diffused across VSe$_2$ layers and then transferred to Gr. The timescale of the interlayer diffusion of hot electrons in MtL VSe$_2$ was much slower than that in 1L VSe$_2$-Gr (~100 fs) (Figure 4a). In the relaxation process of MtL VSe$_2$/Gr, Gr behaves as a cooling channel, and VSe$_2$ behaves as a quasi-static hot electron supply with a long lifetime. These measurements demonstrated the interlayer hot electron transfer in MtL VSe$_2$/Gr heterostructures.

**Coherent acoustic phonon signal in SiC substrate**

During the relaxation process, the TR signal contained a coherent oscillation component. An oscillation frequency of 57 GHz was derived from fast Fourier transform (FFT) results, as shown in Figure 5a. The amplitude of the oscillation component (57 GHz) increased for 1L VSe$_2$/Gr and as the VSe$_2$ layer increased. A clear oscillation signal with the same frequency (57 GHz) was obtained by delivering a higher pump fluence (130 μJ/cm$^2$, Figure 5b). The oscillation signal was launched at a few ps. The oscillation frequency was confirmed by fitting with damped oscillation, including exponential decay. The oscillation frequency did not depend on the thickness of VSe$_2$ layers. This layer–number independence implies that the observed coherent oscillation does not originate from the interlayer oscillation in VSe$_2$ layers. To verify the origin of the oscillation, we performed measurements at different probe wavelengths from 1200 to 1500 nm. Figure 5d shows the measured probe wavelength-dependent oscillation frequency, which is explained by light-matter interaction based on Brillouin scattering.[47-49] The prediction line in Figure 5d, calculated by $f = 2nv_{sound}cos\theta/\lambda_{probe}$, where n is the refractive index of 6H-SiC, $v_{sound} = 13.1$ nm/ps is the sound velocity of 6H-SiC,[51,52] and $\theta = 0°$ is the angle between the pump and probe beam, matched well with the experimental results.

Consequently, we confirmed that the oscillation was related to coherent acoustic phonons propagating through the SiC substrate. Due to the large bandgap (~3 eV) of SiC,[53] photocarriers could not be generated in SiC under 1.5-eV excitation. Accordingly, hot electrons in VSe$_2$ and VSe$_2$/Gr induced the CAP signal in the SiC substrate. As reported in a previous study,[48] the strain pulse was generated by thermal stress from hot phonons during hot electron cooling *via* electron-phonon scattering. In 1L VSe$_2$, hot phonons were generated during the cooling of hot electrons after 2.5 ps; then, the lattice heating induced thermal stress at the VSe$_2$–SiC interface. From this process, the strain pulse propagated and was detectable by the Brillouin backward oscillation of the interference of two reflected probe beams.[49] Gr can also generate thermal stress based on the same principle. The significant CAP signal in Gr on SiC was undetectable (Figure 5a), suggesting that the thermal stress was not sufficient to be measured. The observed CAP signal implies thermal energy transfer from VSe$_2$ to the SiC substrate.

In 1L VSe$_2$/Gr heterostructures, the amplitude of CAP was significantly large upon the formation of heterostructures with Gr (Figure 5a) and proportional to the magnitude of thermal stress.[54] Gr serves as an efficient hot phonon generator through strong coupling with high-energy phonons.[33] The injected hot electrons in 1L VSe$_2$ were transferred to Gr and cooled by emitting hot phonons in Gr, and therefore, the amplitude of CAP increased. The faster decay time in 1L VSe$_2$ with Gr (~38 ps) compared to that without Gr (~200 ps) supports this efficient thermal transfer. Note that Gr acting as an efficient cooling channel and heat spreader was also used in electronic and optoelectronic devices for improving thermal management.[55] Our CAP measurements additionally showed efficient heat energy transfer from $VSe_2$ and heterostructures to the 3D substrate.

**CONCLUSION**

We investigated interlayer coupling and hot electron transfer dynamics in MBE-grown $VSe_2$ and its heterostructures with Gr using time-resolved pump-probe spectroscopy. The combination of the slow dynamics of hot electrons in $VSe_2$ and ultrafast dynamics in Gr accelerated the interlayer hot electron transfer from $VSe_2$ to Gr on a timescale of 100 fs with a very high transfer efficiency (96%). We also demonstrated the hot electron transfer in MtL $VSe_2$/Gr. The observation of CAP showed thermal energy transfer from $VSe_2$ and heterostructures to the 3D substrate.

The observed efficient and ultrafast interlayer hot electron transfer in $VSe_2$/Gr heterostructures is noteworthy and suggests the potential of $VSe_2$ as a layered hot electron injecting material. In vdWHs, a layered structure with sharp 2D interfaces enables efficient interlayer charge transfer for realizing efficient optoelectronic devices.[38-42] So, far, vdWHs have been proposed as semiconducting TMD materials since the available spectral range of injected hot carriers is limited by the bandgap of these materials. Conversely, metals such as gold, which can efficiently generate hot electrons with no spectral limitation, have been used for photo-harvesting,[19] photocatalysts,[56] and nonlinear enhancement.[57] However, since charge transfer competes with ultrafast carrier relaxation, efficient hot electron transfer requires sharp inter-domain[19] or plasmonic coupling.[19,58,59] We expect that a notable category of effective devices may be feasible based on $VSe_2$, *i.e.*, a metallic layer that offers efficient hot carrier injection without spectral restriction as well as sharp 2D interfaces owing to the van der Waals epitaxy.

# METHODS

**Molecular Beam Epitaxy** Monolayer VSe$_2$ was epitaxially grown on Gr deposited on SiC using a home-built ultrahigh-vacuum (UHV) molecular beam epitaxy (MBE) system with a base pressure of 2 × 10$^{-10}$ Torr. We used 6H-SiC (0001) as substrates, supplied by the Crystal Bank at Pusan National University. The SiC substrates were outgassed at 650°C for a few hours and then annealed three times up to 1300°C for 10 s to form Gr, respectively, as verified by reflection high-energy electron diffraction (RHEED) and low-energy electron diffraction (LEED). High-purity V (99.8 %) and Se (99.999%) were simultaneously evaporated by an electron-beam evaporator and a Kundsen cell, respectively, onto the substrate, which was maintained at a temperature of 250°C. The growth process was monitored by *in situ* RHEED, and the growth rate was 5 min per monolayer. After growth, the sample was annealed at 320°C for 30 min and then capped with 1.5-nm Pd at room temperature. Note that the annealing temperature (320°C) is about 60°C lower than the 1T to 2H phase transition temperature of VSe$_2$.[60] For ARPES measurements, we covered the samples with an amorphous selenium layer (100 nm) at room temperature to protect the pristine surface from air exposure after the film growth. The samples were then annealed at 480 K in UHV to remove the selenium capping layer.

**Angle-resolved photoemission spectroscopy.** ARPES measurements were performed in the micro-ARPES end-station (base pressure of ~3×10$^{-11}$ Torr) at the MAESTRO facility at beamline 7.0.2 at the Advanced Light Source, Lawrence Berkeley National Laboratory, using p-polarized photon. The ARPES system was equipped with a Scienta R4000 electron analyzer. The lateral size of the synchrotron beam was estimated to be between 30 and 50 μm. The sample temperature was kept at ~180 K during ARPES measurements. The total energy resolution was 25 meV at $h\nu$ = 115 eV and calibrated with a polycrystalline gold film. The photon energy used was 115 eV.

**Time-resolved two-color pump-probe spectroscopy** Time-resolved reflectivity measurements were performed under ambient conditions at room temperature. A Ti:sapphire laser oscillator (MAITAI, Spectra-Physics), which produces 100-fs pump pulses near 830 nm at an 80-MHz repetition rate was used as an excitation source. The main portion of the output from the oscillator was used to pump a synchronously pumped optical parametric oscillator (SPOPO), which generates tunable probe pulses (1120–1600 nm) with a pulse duration of 150 fs at the same repetition rate. The incident fluence of the pump beam on the sample was adjusted using two polarizers. The time delay between the pump and probe beams was controlled using a motorized translation stage. In transient reflectivity experiments, the pump and probe beams were collinearly combined by a dichroic beam splitter and focused on the sample to a beam waist of 4.5 μm by an objective lens. The pump beam was modulated with a chopper and subsequently blocked by an 800-nm band-pass filter to eliminate the reflected pump signal. The time delay-dependent reflected probe beam was subsequently detected with a lock-in amplifier (SR830, Stanford Research Systems) and a Ge photodetector (DET50B, Thorlabs). Based on this setup, we measured time-resolved differential reflectivity by $\Delta R/R_0 \equiv (R_0 - R'(t))/R_0$, where $R_0$ and $R'(t)$ are the reflectivity of samples without and with pump excitation, respectively.

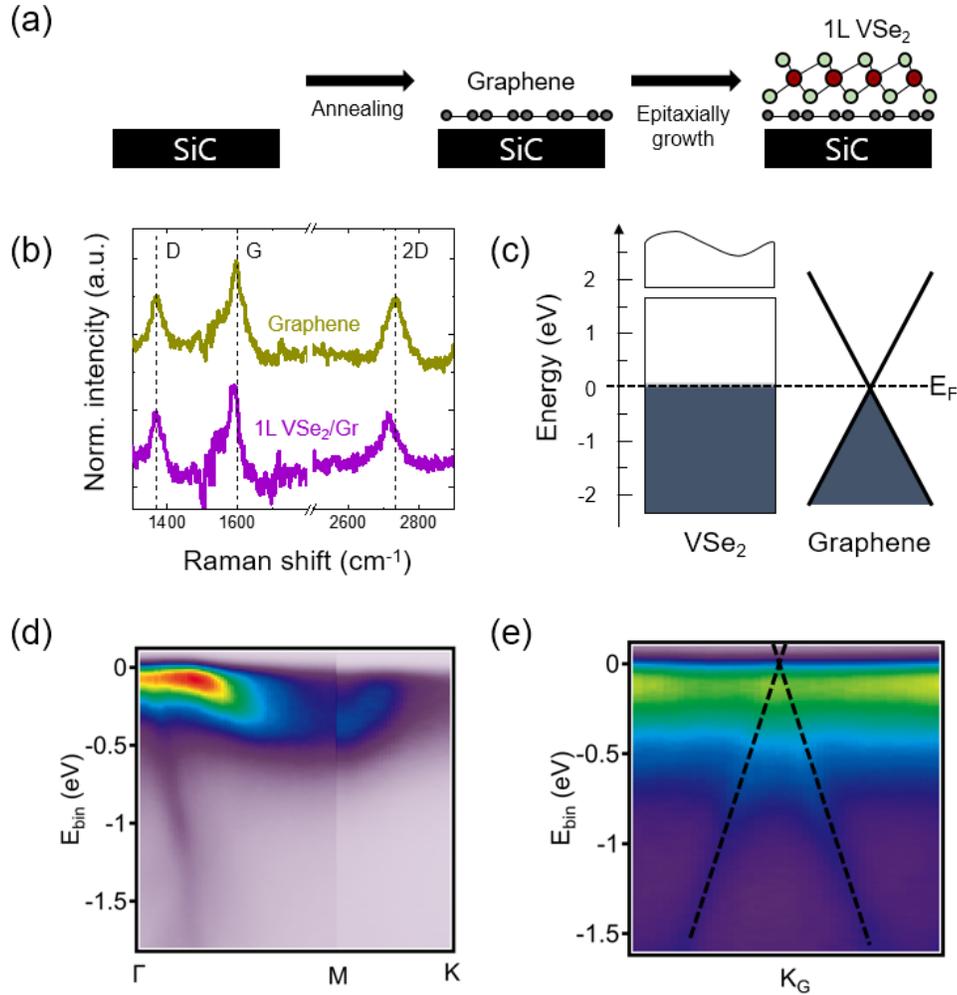

**Figure 1**. Sample preparation and structural and electronic properties of 1L VSe$_2$/Gr heterostructure. **(a)** Fabrication of 1L VSe$_2$/Gr heterostructure on SiC. **(b)** Raman spectra of Gr without and with 1L VSe$_2$. **(c)** Schematic of band alignment in 1L VSe$_2$/Gr heterostructure. The dashed horizontal line indicates the Fermi level in 1L VSe$_2$/Gr. **(d)** ARPES spectrum of 1L VSe$_2$/Gr along the high symmetry direction (Γ-M-K) of 1L VSe$_2$. **(e)** ARPES spectrum of 1L VSe$_2$/Gr near the K point of Gr (K$_G$), emphasizing the linear Dirac band structure of Gr. The dashed lines are guide for the linear band in Gr.

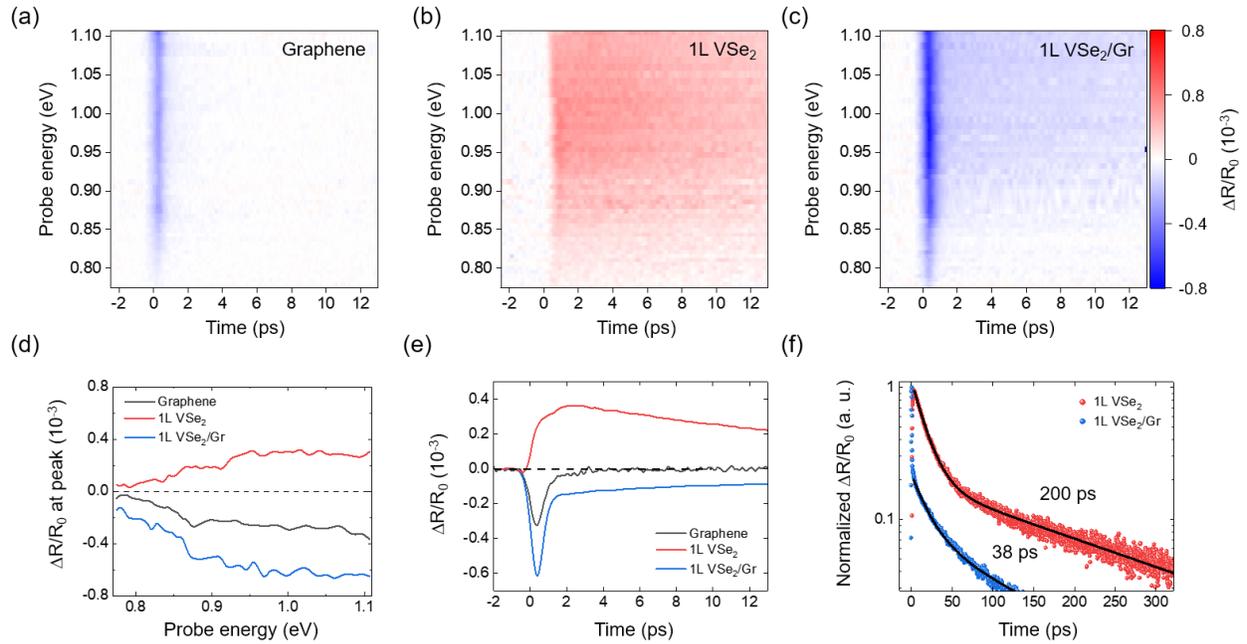

**Figure 2**. Transient reflectivity (TR) measurements at different probe energies under 1.5-eV pump excitation. Two-dimensional plots of TR in **(a)** Gr, **(b)** 1L VSe$_2$, and **(c)** 1L VSe$_2$/Gr. **(d)** Selected TR spectra at peak positions for Gr near 0.2 ps, 1L VSe$_2$/Gr near 0.2 ps, and 1L VSe$_2$ near 2.5 ps. **(e, f)** Time-resolved traces of TR measurements at different time delay ranges. The measured decay times in 1L VSe$_2$ and 1L VSe$_2$/Gr are 200 and 38 ps, respectively.

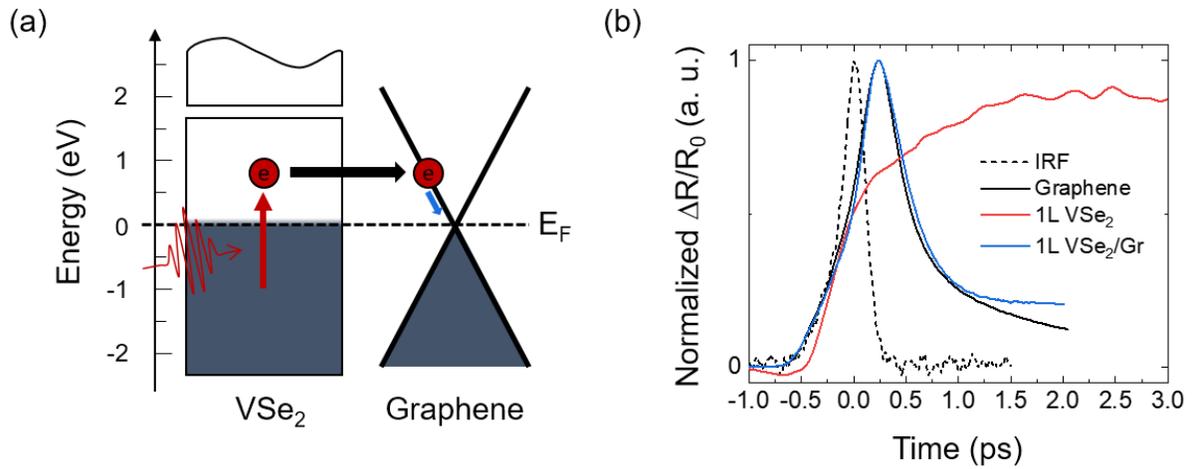

**Figure 3**. Interlayer hot electron transfer at the 1L VSe$_2$/Gr interface. **(a)** Schematic diagram of interlayer hot electron transfer from VSe$_2$ to Gr. The metallic electronic structure of VSe$_2$ and the linear band of Gr are depicted. **(b)** Time-resolved TR results in Gr, 1L VSe$_2$, and 1L VSe$_2$/Gr. The instrumental response function (IRF) indicates a temporal resolution of ~100 fs (dashed line).

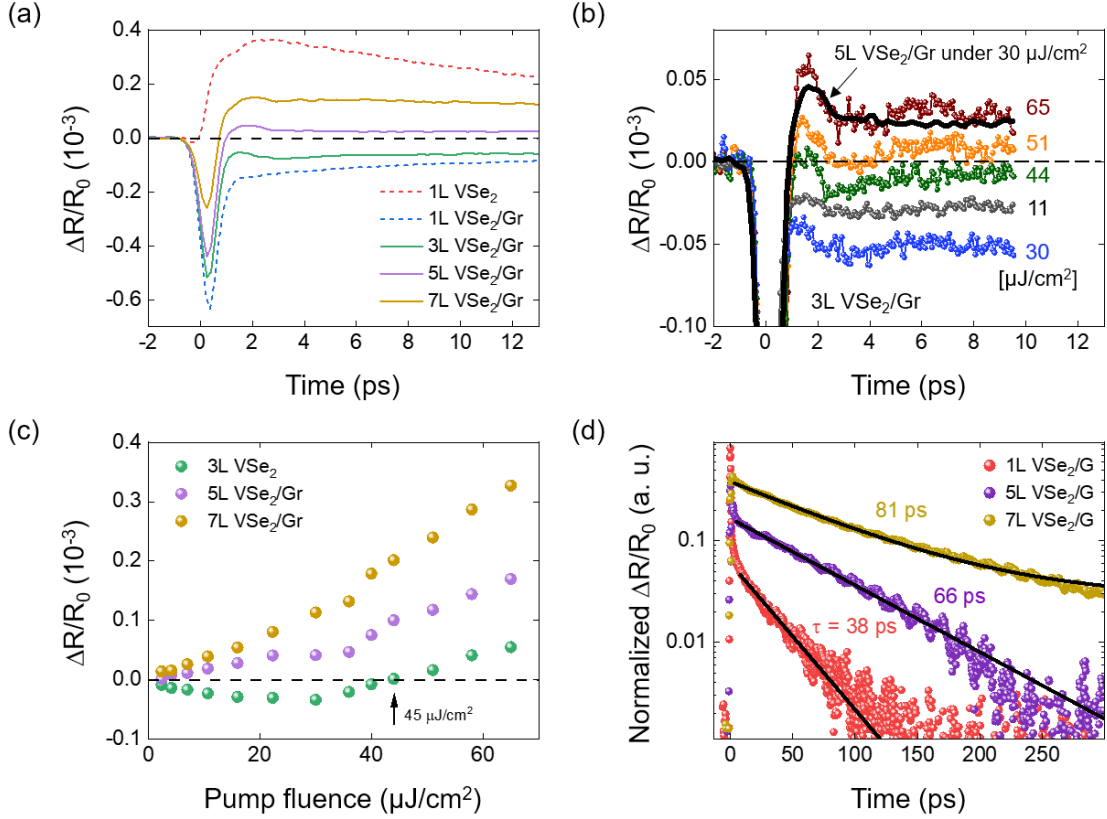

**Figure 4.** Transient reflectivity (TR) of MtL VSe$_2$/Gr heterostructures. **(a)** Time-resolved TR measurements in 3L, 5L, and 7L VSe$_2$/Gr at a pump fluence of 30 μJ/cm$^2$. The TR signals in 1L VSe$_2$ (red dashed line) and 1L VSe$_2$/Gr (blue dashed line) are included for comparison. **(b)** TR signals in 3L VSe$_2$/Gr at various pump fluences. The purple line indicates the TR signal in 5L VSe$_2$/Gr measured at a pump fluence of 30 μJ/cm$^2$. **(c)** Pump-fluence-dependent TR signals obtained from each sample at 1 ps after excitation. **(d)** Decay characteristics in MtL VSe$_2$/Gr. The black solid lines indicate fitting curves with single-exponential decay.

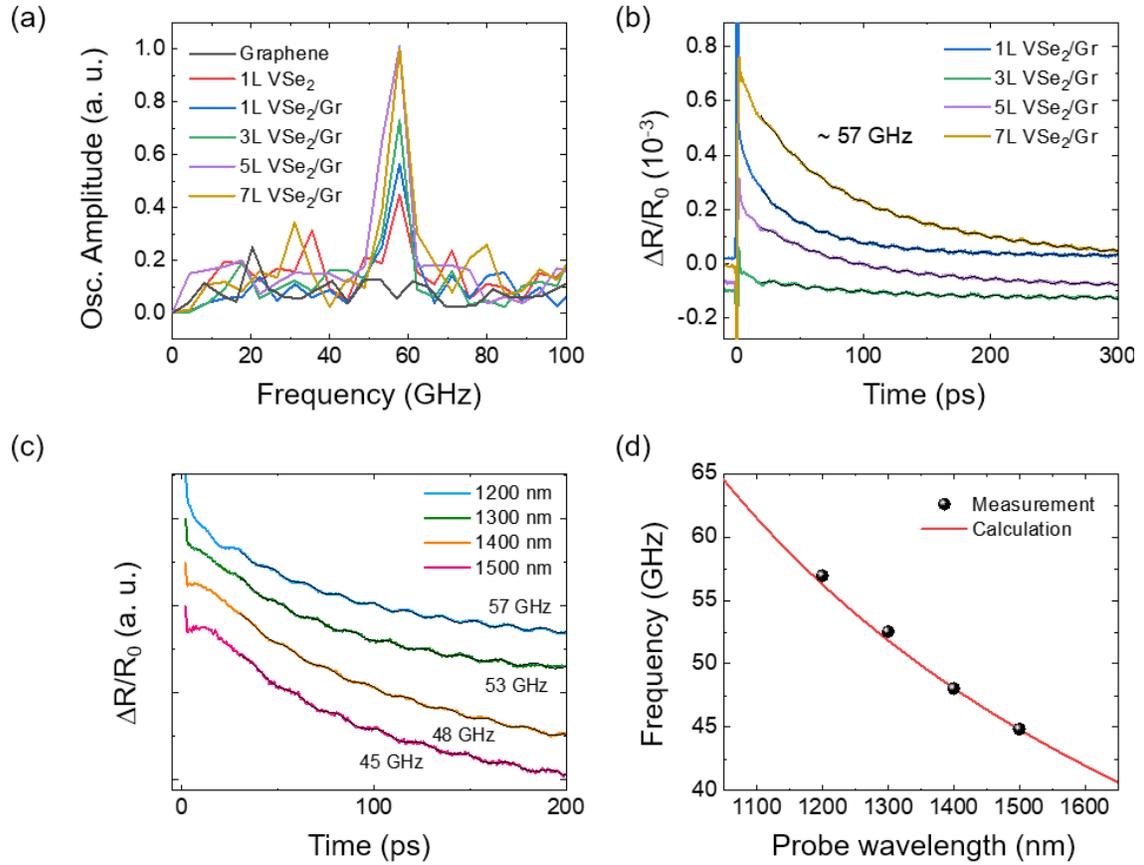

**Figure 5.** Coherent acoustic phonons (CAPs) in VSe$_2$/Gr heterostructures. **(a)** Fast Fourier transform (FFT) results from the TR signal of each sample. **(b)** Enhanced CAP signals measured under a pump fluence of 130 μJ/cm$^2$. **(c)** CAP signals in 5L VSe$_2$/Gr measured at different probe wavelengths between 1200 and 1500 nm. The black fitting curves are obtained based on damped oscillation combined with exponential decay. A period of 17.5 ps corresponding to 57 GHz is listed. **(d)** Measured probe-wavelength-dependent CAP frequencies (black dots). The red curve indicates the calculation result from Brillouin backward scattering.

**ASSOCIATED CONTENT**

The Supporting Information is available free of charge at http://pubs.acs.org/....

Confirmation of influence of Pd capping layer and SiC substrate, pump-fluence-dependent TR measurements in 1L VSe$_2$ and 1L and MtL VSe$_2$/Gr heterostructures, and TR measurements in 1L VSe$_2$ heterostructures on multilayer Gr (PDF)

**AUTHOR INFORMATION**

**Corresponding Author**

**Young Jun Chang** - Department of Physics, University of Seoul, Seoul 02504, Republic of Korea, Department of Smart Cities, University of Seoul, Seoul 02504, Republic of Korea; orcid.org/0000-0001-5538-0643; Email: yjchang@uos.ac.kr

**Fabian Rotermund** - Department of Physics, Korea Advanced Institute of Science and Technology (KAIST), Daejeon 34141, Republic of Korea; orcid.org/0000-0001-9053-3916; Email: rotermund@kaist.ac.kr

**Authors**

**Tae Gwan Park** - Department of Physics, Korea Advanced Institute of Science and Technology (KAIST), Daejeon 34141, Republic of Korea; orcid.org/0000-0002-6158-0091

**Byoung Ki Choi** − Department of Physics, University of Seoul, Seoul 02504, Republic of Korea; orcid.org/0000-0003- 3080-2410

**Junho Park** - Department of Physics, Korea Advanced Institute of Science and Technology (KAIST), Daejeon 34141, Republic of Korea; orcid.org/0000-0002-6955-9457

**Jungdae Kim** - Department of Physics, University of Ulsan, Ulsan 44610, Republic of Korea; orcid.org/0000-0002-8567-1529

**Author Contributions**

†These authors contributed equally. T.G.P and F.R. conceived the original idea. B.K.C. and Y.J.C synthesized and fabricated the VSe$_2$/graphene heterostructure samples. T.G.P. and J.P. performed the experiments. T.G.P. and F.R. and analyzed the data. T.G.P., B.K.C, J.D.K., Y.J.C and F.R. discussed the results. T.G.P. and F.R. wrote the manuscript with input from all authors.

**Funding Sources**

The National Research Foundation of Korea (NRF) funded by the Korean Government (2019R1A2C3003504, 2020R1A4A2002828, 2019K1A3A7A09033389, 2020R1A2C200373211).

**ACKNOWLEDGMENT**

This work was supported by the National Research Foundation of Korea (NRF) funded by the Korean Government (2019R1A2C3003504, 2020R1A4A2002828, 2019K1A3A7A09033389, 2020R1A2C200373211). Authors thank C. Jozwiak, A. Bostwick and E. Rotenberg for ARPES measurements.

56. Kumar, D.; Lee, A.; Lee, T.; Lim, M.; Lim, D.-K., Ultrafast and Efficient Transport of Hot Plasmonic Electrons by Graphene for Pt Free, Highly Efficient Visible-Light Responsive Photocatalyst. *Nano Lett.* **2016**, 16 (3), 1760-1767.

57. Yu, Y.; Si, J.; Yan, L.; Li, M.; Hou, X., Enhanced Nonlinear Absorption and Ultrafast Carrier Dynamics in Graphene/Gold Nanoparticles Nanocomposites. *Carbon* **2019**, 148, 72-79.

58. Brongersma, M. L.; Halas, N. J.; Nordlander, P., Plasmon-Induced Hot Carrier Science and Technology. *Nat. Nanotechnol.* **2015**, 10 (1), 25.

59. Gilbertson, A. M.; Francescato, Y.; Roschuk, T.; Shautsova, V.; Chen, Y.; Sidiropoulos, T. P.; Hong, M.; Giannini, V.; Maier, S. A.; Cohen, L. F., Plasmon-Induced Optical Anisotropy in Hybrid Graphene–Metal Nanoparticle Systems. *Nano Lett.* **2015**, 15 (5), 3458-3464.

60. Li, D.; Wang, X.; Kan, C.-m.; He, D.; Li, Z.; Hao, Q.; Zhao, H.; Wu, C.; Jin, C.; Cui, X., Structural Phase Transition of Multilayer $VSe_2$. *ACS Appl. Mater. Interfaces* **2020**, 12 (22), 25143-25149.

# For Table of Contents Only

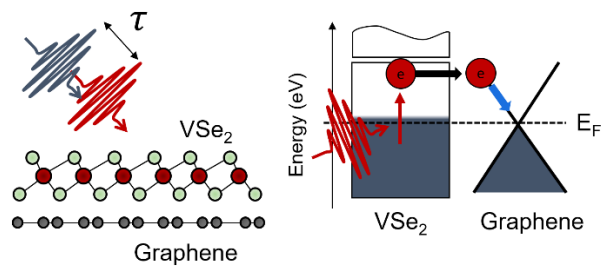

<TOC graphic>

# **Supporting Information**

Interlayer Coupling and Ultrafast Hot Electron Transfer Dynamics in Metallic VSe$_2$/Graphene van der Waals Heterostructures

Tae Gwan Park[1,†], Byoung Ki Choi[2,†], Junho Park[1], Jungdae Kim[3], Young Jun Chang[2,4,*] and Fabian Rotermund[1,*]

[1]Department of Physics, Korea Advanced Institute of Science and Technology (KAIST), Daejeon 34141, Republic of Korea

[2]Department of Physics, University of Seoul, Seoul 02504, Republic of Korea

[3]Department of Physics, University of Ulsan, Ulsan 44610, Republic of Korea

[4]Department of Smart Cities, University of Seoul, Seoul 02504, Republic of Korea

*Correspondence to: yjchang@uos.ac.kr (Y.J.C); rotermund@kaist.ac.kr (F.R.)

†These authors contributed equally to this work.

**Note 1: Confirmation of influence by Pd capping layer and SiC substrate**

Figure S1a shows the results from the time-resolved reflectivity (TR) measurements of the palladium (Pd) capping layer and SiC substrate under the same conditions described in the main manuscript. There is no considerable signal from Pd/SiC because of the large bandgap (>3.2 eV) in 6H-SiC[1] compared to the photon energies of the pump and probe pulses. Figure S1b shows the measured TR signal with or without the Pd capping layer. No significant influence of Pd on the carrier dynamics of Gr was observed. The decay time was almost constant and independent of the existence of the capping layer. The slight decrease in the magnitude of the TR signal is attributed to the absorption or reflection of the pump and probe beams by the capping layer. Consequently, the influences of the Pd capping layer and SiC substrate on dynamic characteristics were neglected in our study.

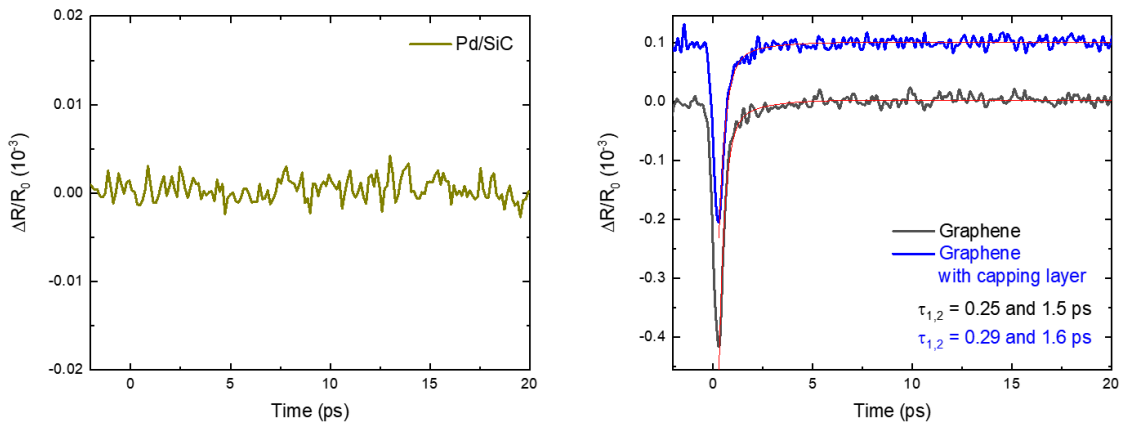

**Figure S1. (a)** Transient reflectivity (TR) of the Pd capping layer on SiC. **(b)** Time-resolved TR of Gr with and without the Pd capping layer at 1.5-eV pump and 1.0-eV probe.

**Note 2: Pump-fluence-dependent TR measurements in 1L VSe$_2$ and 1L and MtL VSe$_2$/Gr heterostructures**

Figure S2 shows the pump-fluence-dependent reflectivity of 1L VSe$_2$. The pump fluence was adjusted to a fluence level below the damage threshold, where the results were highly reproducible. The peak TR values showed a linear dependence on the pump fluence in both 1L VSe$_2$ and 1L VSe$_2$/Gr (Figure S1c), indicating that the nonlinear effect was not involved in the measurements.

Figure S3 shows pump-fluence-dependent decay characteristics. The photoexcited electrons cooled down with two decay components, ~16 and ~200 ps in 1L VSe$_2$. As shown in Figures S2c-d, no significant changes were observed in decay constants, despite almost a 25-fold difference in pump fluence. Figure S4 shows the pump-fluence-dependent TR signals in MtL VSe$_2$/Gr (as used in Figures 4b-c).

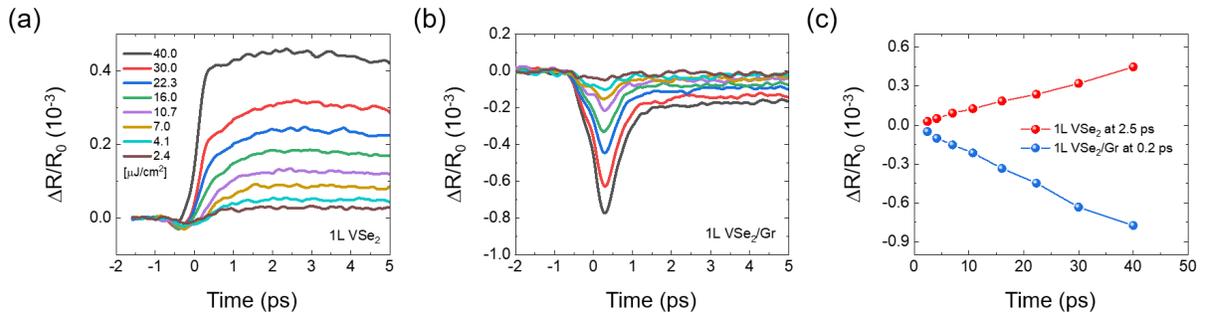

**Figure S2.** Pump-fluence-dependent TR signal in **(a)** 1L VSe$_2$ and **(b)** 1L VSe$_2$/Gr. **(c)** Pump-fluence-dependent peak TR values in 1L VSe$_2$ at 2.5 ps and 1L VSe$_2$/Gr at 0.2 ps.

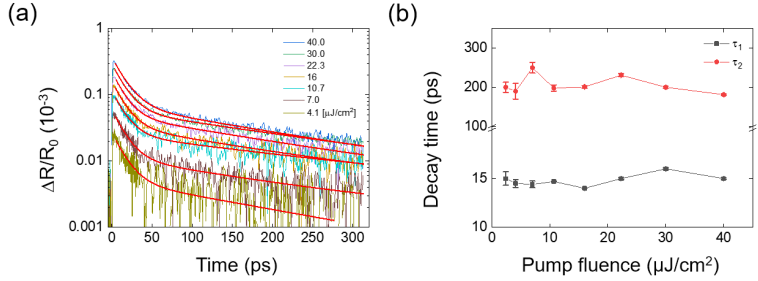

**Figure S3. (a)** Logarithm plots of full-range carrier dynamics in 1L VSe$_2$ at various pump fluences. The red fitting curves indicate the fitting results with bi-exponential decays. **(b)** Pump-fluence-dependent fast (black squares) and slow (red squares) decays in 1L VSe$_2$ estimated by bi-exponential fitting.

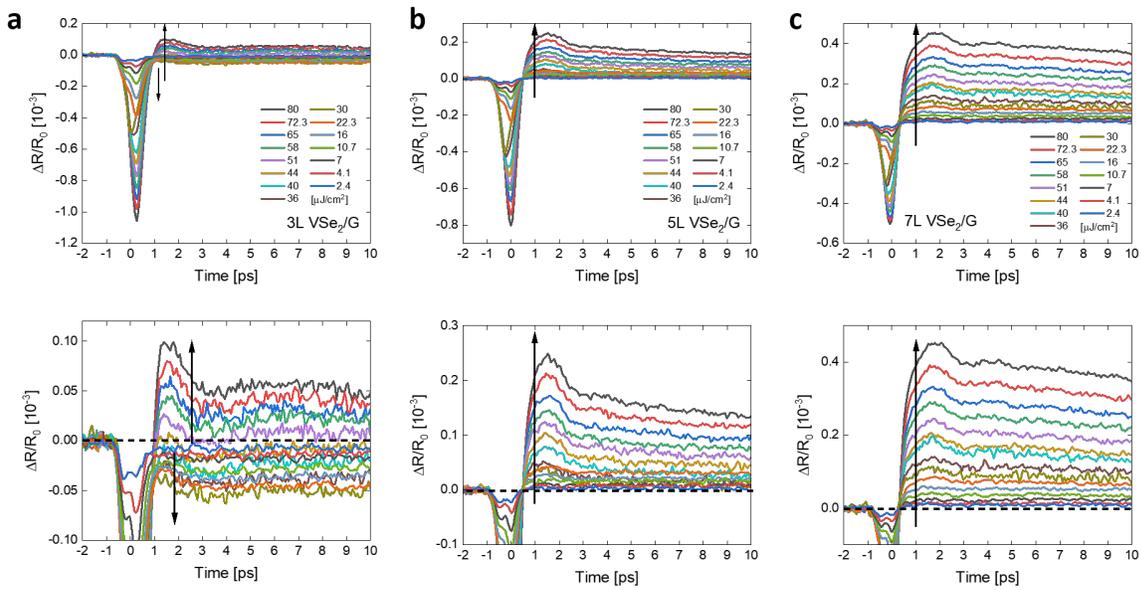

**Figure S4.** Pump-fluence-dependent carrier dynamics in (a) 3L, (b) 5L, and (c) 7L VSe$_2$/G heterostructures. The black arrows indicate the increased pump fluence. The bottom row is magnified for each sample to clearly show the PIA response by localized hot electrons in VSe$_2$ layers. The dashed horizontal lines indicate zero differential reflectivity.

**Note 3: TR measurements in 1L VSe$_2$ heterostructures on multilayer Gr**

The thickness of epitaxial Gr on SiC was controlled by the growth temperature and rate. The multilayer Gr (MLG) was grown for an additional 6 min at 1300 °C, and subsequently, 1L VSe$_2$ was epitaxially grown on this MLG. As shown in Figure S4, we performed the same 1.5-eV pump–1.0-eV probe measurements in 1L VSe$_2$/MLG heterostructures at a pump fluence of 30 μJ/cm$^2$. The photoexcited carrier dynamics in 1L VSe$_2$/MLG were similar to those in 1L VSe$_2$/Gr, as discussed in the main text (Figure S4a). The PIA response in 1L VSe$_2$ was entirely quenched, and only the PB response was dominant by the interlayer hot electron transfer from VSe$_2$. The first decay component (about 0.2 ps) was nearly identical in MLG and 1L VSe$_2$/MLG.

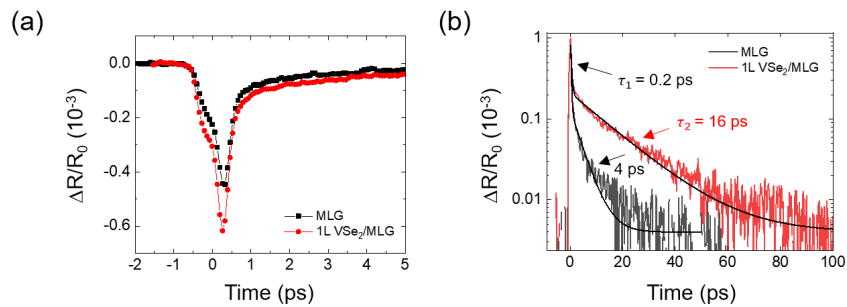

**Figure S5.** Time-resolved traces of TR measurements in multilayer Gr (MLG) and 1L VSe$_2$/MLG in **(a)** short and **(b)** long time delay range.

The slow decay (4 ps) in MLG was slightly slower than that measured in single-layer Gr (Figure 2e). This longer decay time of MLG compared to that of Gr has been reported.[2] Conversely, a decay time of 16 ps in 1L VSe$_2$/MLG was significantly faster than that in 1L VSe$_2$ (~200 ps). Besides, the decay time of MLG was slower than Gr (~1 ps), but the decay time of 1L VSe$_2$/MLG was faster than that of 1L VSe$_2$/Gr (~38 ps). This implies that the MLG provided an abundant cooling channel compared to Gr for cooling the hot electrons of 1L VSe$_2$. This result shows efficient interlayer hot electron transfer from VSe$_2$ to MLG.